\documentclass[11pt]{amsart}

\title{CRYPTOGRAPHIC ANALYSIS OF THE MODIFIED MATRIX MODULAR CRYPTOSYSTEM }

\author{Vitali\u\i\ Roman'kov}
\address{Institute of Mathematics and Information Technologies\\Dostoevsky Omsk State  University}
\curraddr{}
\email{romankov48@mail.ru}

\usepackage{amsmath,amsthm,amsfonts,amssymb}
\usepackage{graphicx}
\usepackage{hyperref}
\usepackage[usenames]{color}

\theoremstyle{definition}

\newcounter{comcount}

\date{}

\begin{document}

\maketitle

\begin{abstract}
We show that the Modified Matrix Modular Cryptosystem  proposed by S.K. Rososhek is not secure against the attack based on the linear decomposition method.  The security of the encryption scheme in the Rososhek’s system is based on the mix of the conjugacy search problem and random "salt"{}.  We do not solve the conjugacy search problem and we are not looking for the exact meaning of the "salt"{}.   The transported secret message in the system is recovered without computation the secret parameters, that have been used for its encryption.   
\end{abstract}

\section{Introduction}
\label{se:intro}
The Basic Matrix Modular Cryptosystem (BMMC) is a public key cryptosystem, which was proposed by S.K. Rososhek in \cite{Ros1}.  Protocol using BMMC was developed for the key exchange in \cite{Ros2}. BMMC realization needs three matrix modular exponentiations for key generation, three exponentiations under encryption and two exponentiations under decryption for every data block. 

In \cite{Ros3}, Rososhek proposed two different modifications of BMMC. We consider them as two versions of the Modified Matrix Modular Cryptosystem (MMMC). The  aim of \cite{Ros3} was to decrease the number of  exponentiations and consequently to accelerate the execution of encryption algorithm. The author of \cite{Ros3} proposed  to determine the large abelian subgroup in general linear group over the large residue ring and to choose the session keys in this subgroup, what will be to give the encryption without exponentiations. Below we consider one of the main protocols, proposed in \cite{Ros3}. 

In this paper, we show that MMMC is vulnerable against the attack based on the linear decomposition method invented by the author in  monograph \cite{Ralg} and papers \cite{Roman}, \cite{MR} with A. Myasnikov (see also monograph \cite{RomEsscrypto}) and developed by the author et al. in the papers  \cite{Rom1} - \cite{Rom4}. In this paper, we describe the attack to MMMC in general case and illustrate the efficiently of this attack on the example of the numerical realization of MMMC proposed by Rososhek in \cite{Ros3}. The security of the encryption scheme in the Rososhek’s system is based on the mix of the conjugacy search problem and random "salt"{}.  We do not solve the conjugacy search problem and we do not seek exact value of the "salt"{}.   The transported secret message is recovered without computation the secret parameters, that have been used for its encryption.   

In \cite{Ralg} (see also \cite{Roman}, \cite{MR} or \cite{RomEsscrypto}), we have shown that in many systems and schemes of the algebraic cryptography, where  the platform group $G$ is a subset of  a linear space, we can efficiently compute the secret message or shared key  and hence to compromise the corresponding cryptographic system. We elaborated a method, that is called the {\it linear decomposition method}. In some points this method is similar to the Tsaban's span method (see \cite{Ts3}). 

The linear decomposition method can be applied if the platform is  part of a finite dimensional linear space $V$ over a field $\mathbb{F},$ or (in the modified form) a finite generated module $M$ over a commutative ring $K.$ In the applications of this method we  construct a basis of the corresponding subspace $W$. In fact, we don't need in a basis. It is sufficient to take a linear generating set for $W$. It is well known that  if $\mathbb{F}$ is a finite field of order $q$ and dim($W$) $=n$, where $n$ is very small with respect to $q$,   then $n$ random vectors of $W$ generate $W$ with high probability. More precisely, let $\mathbb{F}$ be a finite field $\mathbb{F}_q, q = p^r, p $ - prime, of order $q$. Then there are exactly $q^n$ different $n$-vectors over $\mathbb{F}.$ Suppose that we take vectors in random via the uniform distribution on $W$. Let we choose $n$ random vectors in $W$ in sequence. We compute a probability to choose $n$ linearly independent vectors. The first of them is nontrivial with the probability $\frac{g^n-1}{q^n}.$ The second is independent with first with probability $\frac{q^{n} - q}{q^n}, $ and so on till $\frac{q^{n} - q^{n-1}}{q^n}.$ Hence this probability is 
$$\prod_{i=0}^{n-1}\frac{q^n-q^i}{q^n}> (1 - 1/q)^n > 1 - n/q.$$ 
If we know that $W$ has a dimension less than $n$ the corresponding probability to choose  $n$ random elements that generate $W$ is obviously greater than $1-n/q.$

Thus, we can choose randomly elements $w_1, ..., w_n$ in $W$ and try to present the given element $w\in W$ as a linear combination of these elements. We can take more than $n$ elements to increase the mentioned probability. Sure, we need to determine what is "random" in each specific case. Sometimes we can use this approach for modules. Below we'll give the corresponding example. 

Some proposed cryptographic schemes are such that  a linear generating set for $W$ can be easily extracted from the scheme setting. For example, a commutative subsemigroup $G$ of M$_n$($\mathbb{Z}$) can be proposed as follows. Fix a matrix $a \in $  M$_n$($\mathbb{Z}$), and define $G := \mathbb{Z}[a] := \{p(a) : p( t ) \in \mathbb{Z}[t]\}.$ With respect to matrix multiplication $G$ has the structure of  abelian semigroup. Then Alice chooses a matrix $g \in \mathbb{ Z}[a]$ and sends to Bob  vector $hg$, where $h\in H$  is the chosen vector of the protocol.  Bob acts in the similar way. In the cryptanalysis we need to construct a linear generating set for some linear space of the form $vG = \{vg:  g \in G\}$ where $v\in H$. Then in view of the Cayley-Hamilton theorem, the space $vG$ is generated by vectors $v, va, va^2, ... , va^{n-1}.$ In a similar two-sided version where $v$ is a $n\times n$-matrix with rows in $H$, there is a generating set $\{a^iva^j : i, j = 0, ... n-1\}.$ 

\section{ Description of MMMC}
As usual we suppose, that there are two correspondents, Alice and Bob, and that they use a non-secure net for their communications. A potential intruder, Eve, can read all their messages.  

In \cite{Ros3}, Rososhek proposed a cryptographic scheme and considered its a numerical variant. We'll analyze the scheme and this variant and show how the shared common key can be efficiently computed without the secret parameters that has been used for the encryption.
  
\medskip
{\it Assumptions.}
Alice doing the following:
\begin{enumerate}
\item  picks a pair of random prime numbers  $q \not= p$ and computes $n =  pq$, 
then she determines $\mathbb{Z}_n;$
\item takes the obviously abelian subgroup $G= \{\left(\begin{array}{cc}
g&f\\
f&g\end{array} \right) : g,f \in \mathbb{Z}_n$ and $g^2-f^2=1$\};
\item picks four random integers $a, b, c, d \in \mathbb{Z}_n$ such that $a^2-b^2=1$ and $c^2-d^2=1$; 
\item composes two random matrices:
\begin{equation}
\label{eq:41.1}
V = \left(\begin{array}{cc}
a&b\\
b&a\end{array} \right)\in G  \  \textrm{and} \  W=\left(\begin{array}{cc}
c&d\\
d&c\end{array} \right)\in G;
\end{equation}
\item Alice defines two commuting inner automorphisms of the ring M$_2$($\mathbb{Z}_n$):
$\alpha :D \mapsto V^{-1}DV, \beta : D \mapsto W^{-1}DW$
for every matrix $D \in $ M$_2(\mathbb{Z}_n)$.
\item Alice computes the following automorphisms of the ring M$_2(\mathbb{Z}_n)$:
$\psi = \alpha^2\beta , \varphi = \alpha \beta^2.$
\item Alice picks a random invertible matrix $L \in $ GL$_2(\mathbb{Z}_n)$ 
such that $L$ does not belong to the subgroup $G$.
\item  Alice public key is $(n, \varphi (L), \psi (L^{-1})),$
private key is $(V,W).$
\end{enumerate}

\medskip
{\it Algorithm.} 
Bob doing the following:
\begin{enumerate}
\item  presents the plaintext $m$ as a sequence of $2\times 2$-matrices over residue ring $\mathbb{Z}_n$:
$$m_1 || m_2 || ... || m_n;$$
\item for every $m_i, i =1,2,...,n,$ chooses a random matrix $Y_i \in G$;
\item defines for every $i=1,2, ... ,n,$ the automorphisms
$$\xi_i : D \mapsto Y_i^{-1}DY_i \ \textrm{ for every} \   D\in  \  \textrm{M}_2(\mathbb{Z}_n);$$
\item  computes for every $i=1,2, ... , n$ matrices
$$\xi_i(\varphi (L)), \xi_i(\psi (L^{-1})), m_i\xi_i(\varphi (L));$$
\item  picks for every $i=1,2, ... , n$ random units $\gamma_i \in \mathbb{Z}_n^{\ast}$  ("salt") and computes the ciphertext:
$$C = (C^{(1)}||  C^{(2)}||  ... || C^{(n)}),  \   C^{(i)}= (C_1^{(i)}, C_2^{(i)}),$$ 
$$\textrm{where} \ C_1^{(i)} = \gamma_i^{-1}\xi_i(\psi (L^{-1})), C_2^{(i)}= \gamma_im_i\xi_i(\varphi (L)), i = 1, 2, ... , n.$$ 
\end{enumerate}
\medskip
{\it  Decryption.} 
Alice doing the following: 
\begin{enumerate}
\item  computes for every $i=1,2, ... , n,$ using her private key:
$$D_i = \alpha^{-1}(\beta (C_1^{(i)})) = \alpha^{-1}(\beta (\gamma_i^{-1}\xi_i(\psi (L^{-1}))));$$
\item  computes for every $i=1,2, ... ,n$ matrices:
$$C_2^{(i)}D_i = \gamma_im_i\xi_i(\varphi (L))D_i=m_i;$$
\item restores the plaintext $m$ from the matrix sequence $m_1, m_2 ,...,m_n.$
\end{enumerate}

\medskip
\section{Cryptanalysis.}
We are going to show that every $m_i$ can be recovered by any intruder that based only on the public data. It is sufficient to show how we can recover one of the blocks $m_i$.  Denote $m = m_i$, $C_1= C_1^{(i)}, C_2=C_2^{(i)}, \xi = \xi_i, \gamma = \gamma_i, i =1, 2, ..., n.$

Everybody can see the following data: $$n, \varphi (L), \psi (L^{-1}) (\textrm{and so} \  \varphi (L^{-1}), \psi (L)), C_1= \gamma^{-1}\xi (\psi (L^{-1})), C_2 = \gamma m\xi (\varphi (L)).$$ 

It is sufficient to compute  $\gamma^{-1}\xi (\varphi (L^{-1}))$ (to swap  $\psi $ to $\varphi $ in $C_1$).

Let $\tilde{G}$ denotes the abelian subgoup of GL$_2$($\mathbb{Z}_n$) consisting of all matrices of the form $\left(\begin{array}{cc}
a&b\\
b&a\end{array} \right)$ where $a^2-b^2$ is invertible in $\mathbb{Z}_n.$ 
Let  $W$  be the set of all linear combinations of all matrices  of the form $\zeta (\psi (L^{-1})),$ in M$_2(\mathbb{Z}_n),$ where  $\zeta $ is a conjugation by a matrix in  $\tilde{G}.$  

We claim that there exists  a set  {$\zeta_1(\psi (L^{-1})), ..., \zeta_k(\psi (L^{-1})),  \zeta_i \in \tilde{G}, i = 1,2, ..., k; k \leq 4$, for which every matrix in $W$ is a linear combination of these matrices over $\mathbb{Z}_n$.  Below we'll explain this assertion.

For a ring $ R$ and an  $R$-module $M$, the set  $E\subseteq M$ is a {\it basis} for $M$ if:
$E$ is a generating set for $M$ that is to say, every element of $M$ is a finite sum of elements of $ E$ multiplied by coefficients in $R$,  and
$E$ is linearly independent, that is, $\alpha_{1}e_{1} + ... + \alpha_{k}e_{k}=0$ for $e_{1}, ... , e_{k} $ distinct elements of $ E$ implies that  $\alpha_{1} = ... = \alpha_{k}=0.$ A {\it free module} is a module with a basis. But not each module has a basis.
 
For any submodule $V$ of the free module $\mathbb{Z}_n^r,$ where $ r \in \mathbb{N}$ and 
$n=pq$,  as above, we define a notion of  a {\it quasi-basis} as a minimal subset  $E$ of $V$ such that every element of $V$ is a finite sum of elements of $E$ multiplied by coefficients in $\mathbb{Z}_n.$ 
Now we prove that $V$ has a quasi-basis consisting of $\leq r$ elements and show how such  quasi-basis  can be obtained.

 Let $V_p$ be the $p$-image of $V$, i.e., a homomorphic image of $V$ modulo $p$ and $V_q$ is the $q$-image of $V$ modulo $q$. Then $V_p$ is a linear space over $\mathbb{Z}_p$, and $V_q$ is a linear space over $\mathbb{Z}_q.$ Let $a_1, ..., a_k$ be a basis of $V_p$ and $b_1, ..., b_l$ be a basis of $V_q$. Since $V_p$ and $V_q$ are subspaces of $\mathbb{Z}_p^r$ and $\mathbb{Z}_q^r$ respectively, $k, l \leq r.$ Suppose that $k \geq l$. If $k\not= l$ we add to the set $b_1, ..., b_l$ $k-l$ zero elements and get $b_1, ..., b_k$.   We can consider elements  $a_i$ and $b_j$ as $r$-tuples of components $a_i^{(j)}$ and $b_i^{(j)}$ that are written as integers. Then by the Chinese remainder theorem we can find $e_i^{(j)} \in \mathbb{N}$ such that  $e_i^{(j)}=a_i^{(j)}(mod \  p)$ and $e_i^{(j)}=b_i^{(j)}(mod \  q),$ respectively. We do it for all $i$ and $j.$ As result we have a quasi-basis $E = \{e_1, ..., e_k\}.$ Indeed, two the images $v_p\in V_p$ and $v_q\in V_q$ of an arbitrary element $v\in V$  have two presentations:
$$v_p= \sum_{i=1}^k\alpha_i e_i   \textrm{ and}  \    v_q= \sum_{i=1}^k\beta_i e_i, \  \textrm{respectively},$$ 
\noindent
where all coefficients are written as natural numbers. Again, by the Chinese remainder theorem we can find $\gamma_i$ such that $\gamma_i=\alpha_i (mod \ p)$ and 
$\gamma_i=\beta_i (mod \ q)$ for each $i=1, ..., k.$ Then 
$$v= \sum_{i=1}^k \gamma_i e_i$$
\noindent is a presentation of $v$ as a linear combination of vectors of  $E$ over $\mathbb{Z}_n.$ Obviously the size $k$ is minimal for a generating set, thus $E$ is quasi-basis.  

The just described algorithm can be applied only in the case when $p$ and $q$ are known.  In other case we only know that there is a quasi-basis consisting of $k\leq r$ elements. 

Now we return to the considering protocol. We have M$_2$($\mathbb{Z}_n$) that is a free $\mathbb{Z}_n$-module of dimension $4,$ and its submodule $W.$ We have just proved that there is a quasi-basis $\{\zeta_1(\psi (L^{-1})), ..., \zeta_k(\psi (L^{-1})),  \zeta_i \in \tilde{G}, i = 1,2, ..., k; k \leq 4\}$ of $W$.  We need not exactly in quasi-basis but in some generating set. 
A set of four such elements chosen by  the random process via uniform distribution generates $W$ if and only if it generates $W$ modulo $p$ and $q$ simultaneously. The corresponding probabilities for $n<p,q$ (as it has been showed above) exceed $1-n/p$ and $1-n/q$ respectively. It follows that the probability to generate $W$ exceeds $(1-n/p)(1-n/q).$ Suppose that we find a generating set $\{\zeta_1(\psi (L^{-1})), ..., \zeta_4(\psi (L^{-1})),  \zeta_i \in \tilde{G}, i = 1, ..., 4\}$ of $W$.

Then we compute a presentation of the form 
\begin{equation}
\label{eq:1.2}
\varphi (L^{-1}) = \sum_{i=1}^4\alpha_i\zeta_i(\psi (L^{-1})), \alpha_i \in \mathbb{Z}_n.
\end{equation}
Then we change  in the right hand  side of (\ref{eq:1.2})  $\psi (L^{-1})$ by $C_1$:

\begin{equation}
\label{eq:1.3}
\sum_{i=1}^4\alpha_i\zeta_i(C_1)=\gamma^{-1}\xi (\sum_{i=1}^4\alpha_i\zeta_i(\psi (L^{-1})) = \gamma^{-1}\xi (\varphi (L^{-1})).
\end{equation}

Now we recover the message as
\begin{equation}
\label{eq:1.4}
C_2 \gamma^{-1}\xi (\varphi (L^{-1})) = m.
\end{equation}

\section{ Example} 

Now we consider the numerical Example 1 in   \cite{Ros3}  and give a cryptanalysis. 

\medskip
{\it Assumptions.}

Alice doing the following: 
\begin{enumerate}
\item picks the primes $p=5, q=7$ and computes $n = pq = 35;$
\item chooses four random integers in the modular ring $\mathbb{Z}_{35}$: $7, 4, 6, 2;$
\item composes the random matrices
$$V = \left(\begin{array}{cc}7&4\\
4&7\end{array}\right), W =\left(\begin{array}{cc}6&2\\
2&6\end{array}\right);$$
\item computes det($V$) = $33$,  det($W$) = $32$ and then computes det($V$)$^{-1}$ = $17$, det($W$)$^{-1}$=$23.$ therefore $V$ and $W$ are units in the matrix ring M$_2$($\mathbb{Z}_{35}$);
\item defines two automorphisms of the ring M$_2$($\mathbb{Z}_{35}$):
$$\alpha : D \mapsto V^{-1}DV, \beta : D \mapsto W^{-1}DW$$ 
\noindent  for every matrix $D \in $ M$_2$($\mathbb{Z}_{35}$);
\item computes the following automorphisms :
$$\psi = \alpha^2\beta , \varphi = \alpha \beta^2;$$ 
\item chooses the random matrix $L \in $ GL$_2$($\mathbb{Z}_{35}$):
$$L = \left(\begin{array}{cc}1&2\\
3&5\end{array}\right)$$
\noindent and computes matrix 
$$L^{-1} = \left(\begin{array}{cc}30&2\\
3&34\end{array}\right);$$
\item computes matrices:
$$\varphi (L) = (VW^2)^{-1}L(VW^2) = \left(\begin{array}{cc}34&34\\
6&7\end{array}\right),$$ 
$$\psi (L^{-1}) = (V^2W)^{-1}L^{-1}(V^2W) = \left(\begin{array}{cc}23&24\\
16&6\end{array}\right);$$
\item Alice public key is 
$$(n=35, \varphi (L) =  \left(\begin{array}{cc}34&34\\
6&7\end{array}\right),  \psi (L^{-1}) =  \left(\begin{array}{cc}23&24\\
16&6\end{array}\right)),$$
\noindent
private key is 
$$(V = \left(\begin{array}{cc}7&4\\
4&7\end{array}\right), W =\left(\begin{array}{cc}6&2\\
2&6\end{array}\right)).$$
\end{enumerate} 

\medskip
{\it Algorithm.}

Bob doing the following:

\begin{enumerate}
\item
presents the plaintext as a matrix
$$m = \left(\begin{array}{cc}11&2\\
9&3\end{array}\right) \in \textrm{M}_2(\mathbb{Z}_{35});$$
\item 
picks the random matrix 
$$Y =  \left(\begin{array}{cc}3&5\\
5&3\end{array}\right) \in G$$
\noindent
and computes 
$$Y^{-1} =  \left(\begin{array}{cc}2&20\\
20&2\end{array}\right);$$
\noindent
defines automorphism $\xi $ of the ring M$_2(\mathbb{Z}_{35})$:
$$\xi : D \mapsto Y^{-1}DY$$
\noindent
for every $D \in $ M$_2(\mathbb{Z}_{35});$
\item computes matrices: 
$$\xi (\varphi (L)) = Y^{-1}(\varphi (L))Y =  \left(\begin{array}{cc}29&24\\
16&12\end{array}\right),$$
$$\xi (\psi (L^{-1}))=Y ^{-1}(\psi (L^{-1}))Y = \left(\begin{array}{cc}13&24\\
16&16\end{array}\right);$$
\item 
picks random unit $$\gamma \in \mathbb{Z}_{35}, \gamma = 9, \gamma^{-1}= 4;$$
\item
computes the ciphertext $C=(C_1, C_2):$
$$C_1 =\gamma^{-1}\xi (\psi (L^{-1})) = \left(\begin{array}{cc}17&26\\
29&29\end{array}\right),$$
$$C_2 = \gamma m \xi (\varphi (L)) =  \left(\begin{array}{cc}9&2\\
16&28\end{array}\right).$$
\end{enumerate}

\medskip
{\it Decryption.}

Alice doing the following:

\begin{enumerate}
\item computes matrix $z,$ using her private key:
$$z = \alpha^{-1}\beta (C_1) =  \left(\begin{array}{cc}22&26\\
29&24\end{array}\right);$$
\item computes then 
$$C_2z =  \left(\begin{array}{cc}11&2\\
9&3\end{array}\right)=m.$$ 
\end{enumerate}

\medskip
{\it Cryptanalysis.} 
Firstly we compute 
$$\varphi (L^{-1}) = (\varphi (L))^{-1}= \left(\begin{array}{cc}
28&34\\
6&1\end{array}\right).$$
By the way we can compute $(\gamma^{-1})^2$. Indeed, det$(\psi (L^{-1})) = 34,$ det$(C_1)=$ det$(\psi (L^{-1}))= 19,$ then $(\gamma^{-1})^2=16.$

We choose four "random" matrices   
 of the form $\zeta (\psi (L^{-1})),$ where $\zeta \in \tilde{G}$ (in fact four matrices with simplest conjugators):
$$
e_1 = \psi (L^{-1}) = \left(\begin{array}{cc}23&24\\
16&6\end{array}\right),$$
$$ e_2 = (-1)\left(\begin{array}{cc}0&34\\
34&0\end{array}\right)\left(\begin{array}{cc}23&24\\
16&6\end{array}\right)\left(\begin{array}{cc}0&1\\
1&0\end{array}\right)=\left(\begin{array}{cc}6&16\\
24&23\end{array}\right),$$
$$
e_3= 12 \left(\begin{array}{cc}2&1\\
1&2\end{array}\right)\left(\begin{array}{cc}23&24\\
16&6\end{array}\right)\left(\begin{array}{cc}2&34\\
34&2\end{array}\right)= \left(\begin{array}{cc}0&27\\
13&29\end{array}\right),$$
\begin{equation}
\label{eq:1.6}
e_4 = 23\left(\begin{array}{cc}1&2\\
2&1\end{array}\right)\left(\begin{array}{cc}23&24\\
16&6\end{array}\right)\left(\begin{array}{cc}1&33\\
33&1\end{array}\right)= \left(\begin{array}{cc}29&13\\
27&0\end{array}\right).
\end{equation} 

Then we are to solve the  equation
$$\varphi (L^{-1}) = \sum_{i=1}^4\alpha_i e_i,$$
\noindent
namely:
$$ \left(\begin{array}{cc}28&34\\
6&1\end{array}\right)=\alpha_1  \left(\begin{array}{cc}23&24\\
16&6\end{array}\right)+\alpha_2  \left(\begin{array}{cc}6&16\\
24&23\end{array}\right) + $$
$$\alpha_3  \left(\begin{array}{cc}0&27\\
13&29\end{array}\right) + \alpha_4   \left(\begin{array}{cc}29&13\\
27&0\end{array}\right).$$

By direct computation via the Gauss elimination process we obtain the unique solution:
$$\alpha_1= 7, \alpha_2 = 0, \alpha_3 = 1, \alpha_4 = 28.$$

We have
$$ \left(\begin{array}{cc}28&34\\
6&1\end{array}\right)= 7 \left(\begin{array}{cc}23&24\\
16&6\end{array}\right) + $$
$$12 \left(\begin{array}{cc}2&1\\
1&2\end{array}\right)\left(\begin{array}{cc}23&24\\
16&6\end{array}\right)\left(\begin{array}{cc}2&34\\
34&2\end{array}\right) + $$
$$  14 \left(\begin{array}{cc}1&2\\
2&1\end{array}\right)\left(\begin{array}{cc}23&24\\
16&6\end{array}\right)\left(\begin{array}{cc}1&33\\
33&1\end{array}\right).$$

Then we swap $$\psi (L^{-1})=\left(\begin{array}{cc}23&24\\
16&6\end{array}\right)$$
\noindent 
 in the right hand side of the last equality with $$C_1=\left(\begin{array}{cc}17&26\\
29&29\end{array}\right)$$
\noindent
 and compute

$$7\left(\begin{array}{cc}17&26\\
29&29\end{array}\right) + 12 \left(\begin{array}{cc}2&1\\
1&2\end{array}\right)\left(\begin{array}{cc}17&26\\
29&29\end{array}\right)\left(\begin{array}{cc}2&34\\
34&2\end{array}\right) +$$
$$14\left(\begin{array}{cc}1&2\\
2&1\end{array}\right)\left(\begin{array}{cc}17&26\\
29&29\end{array}\right)\left(\begin{array}{cc}1&33\\
33&1\end{array}\right) =$$
$$\left(\begin{array}{cc}14&7\\
28&28\end{array}\right)+ \left(\begin{array}{cc}15&33\\
22&31\end{array}\right)+ \left(\begin{array}{cc}28&21\\
14&0\end{array}\right)=
\left(\begin{array}{cc}22&26\\
29&24\end{array}\right). $$
At last we multiply $C_2$ to the just computed matrix:
$$\left(\begin{array}{cc}9&2\\
16&28\end{array}\right)\left(\begin{array}{cc}22&26\\
29&24\end{array}\right)= \left(\begin{array}{cc}11&2\\
9&3\end{array}\right) = m,$$
\noindent 
and we succeeded.

\end{document}